\documentclass[preprint,12pt]{elsarticle}
\usepackage{graphicx}
\usepackage{amssymb}
\usepackage{amsmath}

\journal{Nuclear Physics A}

\begin{document}

\begin{frontmatter}

\title{Feasibility guidelines for kaonic-atom experiments with 
ultra-high-resolution X-ray spectrometry}

\author[a]{E.~Friedman}\corref{cor1}
\cortext[cor1]{Corresponding author: E. Friedman, elifried@cc.huji.ac.il}
\author[b]{S.~Okada}
\address[a]{Racah Institute of Physics, The Hebrew University, 91904
Jerusalem, Israel}
\address[b]{RIKEN Nishina Center, RIKEN, Wako, 351-0198, Japan}
\date{\today}
\begin{abstract}
Recent studies of strong interaction effects in kaonic atoms suggest
that analysing so-called `lower' and `upper' levels in the same 
atom could separate one-nucleon absorption from multinucleon
processes. The present work examines the  feasibility of direct 
measurements of
upper level widths in addition to lower level widths in future experiments, 
using superconducting microcalorimeter detectors.
About ten elements are identified as possible candidates for such experiments,
all of medium-weight and heavy nuclei. New experiments focused on
achieving good accuracy for widths of such pairs of levels could contribute
significantly to our knowledge of the $K^-$-nucleon interaction in the
nuclear medium.

\end{abstract}

\end{frontmatter}

\section{Introduction}
\label{sec:intro}

Results of the precision measurements of kaonic hydrogen atoms by the
SIDDHARTA collaboration \cite{SID11,SID12} form already part of
the data-base used  by Ikeda, Hyodo and Weise (IHW)
in constructing the antikaon-nucleon scattering
amplitudes near threshold \cite{IHW11,IHW12}.
Studies of kaonic atoms using potentials built on sub-threshold
in-medium antikaon-nucleon scattering amplitudes \cite{CFG11a,CFG11b,FGa12}
clearly indicate that multinucleon processes contribute significantly
to the observed strong-interaction level shifts and widths. In particular
it was shown \cite{FGa13} using the IHW amplitudes
that analysing so-called `lower' and `upper' levels 
in the same atom could separate one-nucleon (1N) absorption from multinucleon
(mN) processes. This property is the result of the very different radial
dependences of the 1N and 2N terms of the potential, as demonstrated for
Ni and Pb in table \ref{tab:radii}. It is seen that in both examples
the rms radius of the 1N real term is larger than that for the mN real
term by 0.95 fm and that for the imaginary part the difference is 1.2 to
1.3 fm.

With the one-nucleon amplitudes firmly based on the SIDDHARTA
experiment and its subsequent analyses, there is now a possibility to gain
information on multinucleon processes of antikaons in nuclei. This calls
for reduced uncertainties in experimental results, particularly for
the upper level widths.

\begin{table}
\caption{rms radii of various terms of the K$^-$-nucleus potential (in fm).
r$_m$ is the rms radius of the nucleus.}
\label{tab:radii}
\begin{center}
\begin{tabular}{cccccccc}
\hline
 & r$_m$&Re(full)&Re(1N)&Re(mN)&Im(full)&Im(1N)&Im(mN) \\ \hline
Ni &3.72&3.34&3.82&2.86&3.73&4.46&3.12 \\
Pb &5.56&5.21&5.71&4.78&5.46&6.23&5.00 \\ \hline
\end{tabular}
\end{center}
\end{table}

Strong interaction effects in exotic atoms have been studied in
great detail for several decades, see \cite{FGa07} for a recent
review. Regarding strengths of absorption, kaonic atoms are 
intermediate between weak absorption in pionic atoms and very 
strong absorption in antiprotonic atoms. 
Absorption is sufficiently strong to make it the dominant effect
in kaonic atoms, where strong-interaction level widths are up to 
one order of magnitude larger than the corresponding strong-interaction 
level shifts. Furthermore, these shifts
are almost universally repulsive, although the real potentials required
to fit kaonic atom data are attractive. Hence the role of the real
part is secondary to that of the imaginary part of the potential.

The level width which is 
usually obtained as the imaginary part of the complex eigenvalue 
when solving the Klein-Gordon equation with an optical potential
\cite{BFG97} is also related to the imaginary part of the potential
as follows

\begin{equation}
\label{eq:width}
\Gamma_{\rm st}~=-2~\frac{\int {\rm Im}V_{\rm opt}~|\psi|^2~d{\vec r}}
{\int [1-(B+V_{\rm C})/\mu]~|\psi|^2~d{\vec r}}
\end{equation}
where $B$, $V_{\rm C}$ and $\mu$ are the $K^-$ binding energy, Coulomb
potential and reduced mass, respectively. (For a Schroedinger
equation the denominator is just the normalizing integral.)
The widths are therefore the quantities which are more directly
connected to the potential.

All available experimental results regarding upper levels in 
kaonic atoms are in the form of relative yields, defined as

\begin{equation}
\label{eq:yield}
Y^{\rm rel}~=~\Gamma_{\rm rad}~/~(\Gamma_{\rm rad}~+~\Gamma_{\rm st})
\end{equation}
where $\Gamma_{\rm rad}$ and $\Gamma_{\rm st}$ are the radiation
width of the upper to lower level transition and the strong interaction width
of the upper level, respectively. The relative error of the derived
$\Gamma_{\rm st}$ which is the quantity of interest turns out to be 
the relative error of the measured
yield divided by $(1-Y^{\rm rel})$, introducing conflicting demands between 
intensity and accuracy.

The available data for kaonic atoms are based on experiments of
three to four decades ago \cite{FGB94}. Although the data cover the
whole of the
periodic table with reasonably good accuracy for the shifts and
widths of the lower levels, the widths of the upper levels are
determined from the measured relative yields of the upper to lower
level transitions. This causes an increase of errors on top of
the usually poorer accuracies in the determination of yields. It is
therefore desirable to carry out new experiments on several carefully
selected targets where improved accuracy in determining strong-interaction
observable for both lower and upper levels in the same atom is the prime
concern. Recent development of a high resolution gamma-ray spectrometer
based on superconducting microcalorimeters makes it possible,
for the first time, to consider direct measurements of widths of upper levels
in kaonic atoms.

Section \ref{sec:det} outlines basic features of such high-resolution 
detector and
section \ref{sec:method} explains the method used in the present survey,
looking for suitable targets for future kaonic atom experiments.
All calculations are based as much as possible on information 
gained from global fits to existing data from kaonic atoms.
The results of the survey are presented in section \ref{sec:results}.
These include widths of the various levels involved and 
values of E$_{\rm X}$, the
energy of the X-ray transition feeding into the upper level.
Relative and absolute yields of relevant transitions are also
shown, to serve as guidance for further 
detailed planning of experiments.
Section \ref{sec:disc} is a discussion and summary, addressing 
also the question of the ability of restricted data to provide 
useful information.

\section{Detectors}
\label{sec:det}

In recent years, cryogenic particle detectors have achieved remarkable
development \cite{Enss2005}.
Above all, superconducting microcalorimeters based on
transition-edge sensor (TES) offer unprecedented energy resolution
for single photon detections from the near infrared through gamma rays.
Thanks to recent technological advances in multiplexed readout of
a TES multi-pixel array, the large-area ultra-high-resolution TES
detector has now edged closer to practical use in a variety of nuclear
measurements.

For gamma-ray spectroscopy, the energy resolution for a single pixel
so far achieved is 22-eV resolution (FWHM) at $\sim$100 keV
\cite{Bacrania2009}, being more than an order of magnitude better
than high-purity germanium detectors.
An average resolution of 53 eV at $\sim$100 keV was achieved
for a 256 pixel spectrometer with a collecting area
of 5 cm$^2$ \cite{Bennett2012}.
The useable dynamic range of this detector is above 400 keV
and the energy resolution $\Delta E$ is almost independent of the
X-ray energy within the linear range of the detector. In general
$\Delta E$ is expected to be proportional to the square-root of
this upper limit.

In the present work we take the energy resolution of 53 eV
at $\sim$100 keV and either keep it fixed for all X-ray energies
or scale it for energy $E_X$ keV as $53 \times \sqrt{E_X/100}$ eV.
We chose also this scaling to cover possible different optimizations of
resolution and linearity, to include also more conservative estimates
in the feasibility study.

\section{Method}
\label{sec:method}

In order to base the present study as much as possible on  the
available kaonic atom data we use  phenomenological $K^-$-nucleus
potentials that produces very good fits to the data \cite{FGa13,FGB94},

\begin{equation}
V_{\rm opt}~=~-(4\pi/2\mu )[b\rho+B\rho(\rho/\rho_0)^{0.25}] \; ,
\label{eq:phen}
\end{equation}
where $\mu$ is the kaon-nucleus reduced mass, $b$ and $B$ are
energy-independent complex parameters and  
$\rho$ is the local nuclear density. 
 Parameter values were $b=-0.15+{\rm i}0.62$~fm,
$B=1.65-{\rm i}0.045$~fm, $\rho_0$=0.17 fm$^{-3}$, producing $\chi ^2$=103
for 65 data points. This potential was used 
for calculating binding energies
and strong interaction widths for all kaonic atoms concerned.
Similar results are obtained also from potentials of ref. \cite{FGa13}.

Searching for possible candidates for new kaonic atom experiments,
we are interested in X-ray transitions between three
`circular' states, i.e. where the radial number $n$ is $l+1$.
The lower level (L) is the one where the X-ray cascade terminates due to
dominance of the absorption and the upper level (U) has $n$ larger by 1. 
If the width of the upper level ($\Gamma _{\rm U}$) is to be measured directly, 
then the transition
from yet another level (U+1) must be considered. 
The energy of this transition is denoted by E$_{\rm X}$. Transitions from
larger $n$-values could be useful too.

With the parameters listed above it is possible to calculate strong 
interaction level shifts and widths as well as upper level 
{\it relative} yields
and, of course, also transition energies. However, to complete the
picture regarding feasibility of measurements it is necessary to
obtain also reliable estimate of {\it absolute} yields of the
various transitions. That was achieved by performing calculations
of the atomic cascade process within the exotic atom using a program
due to Batty \cite{Bat99} which was developed from the program
of Huefner~\cite{Hue66}.
Cascade calculations require an initial distribution of the population
of high-$n$ states for the various values of the orbital angular
momentum $l$. A statistical $(2l+1)$ distribution was assumed but
deviations from it were also considered by multiplying $(2l+1)$ 
by $e^{\alpha l}$ for several values of $\alpha$ 
as was done in other studies of exotic atoms \cite{BFG98}.

\section{Results}
\label{sec:results}

\begin{figure}[htb]
\begin{center}
\includegraphics[height=95mm,width=0.45\textwidth]{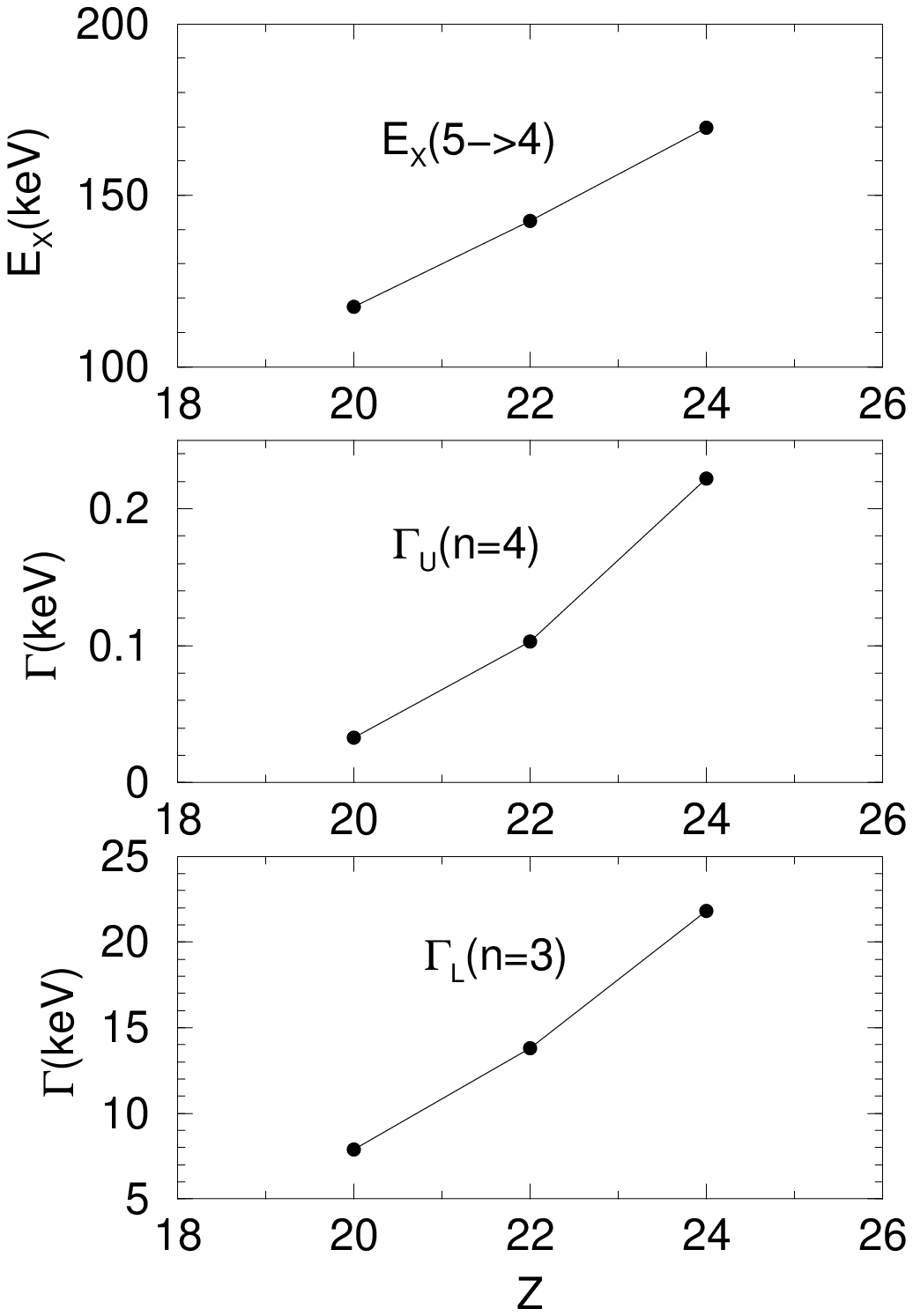}
\includegraphics[height=95mm,width=0.45\textwidth]{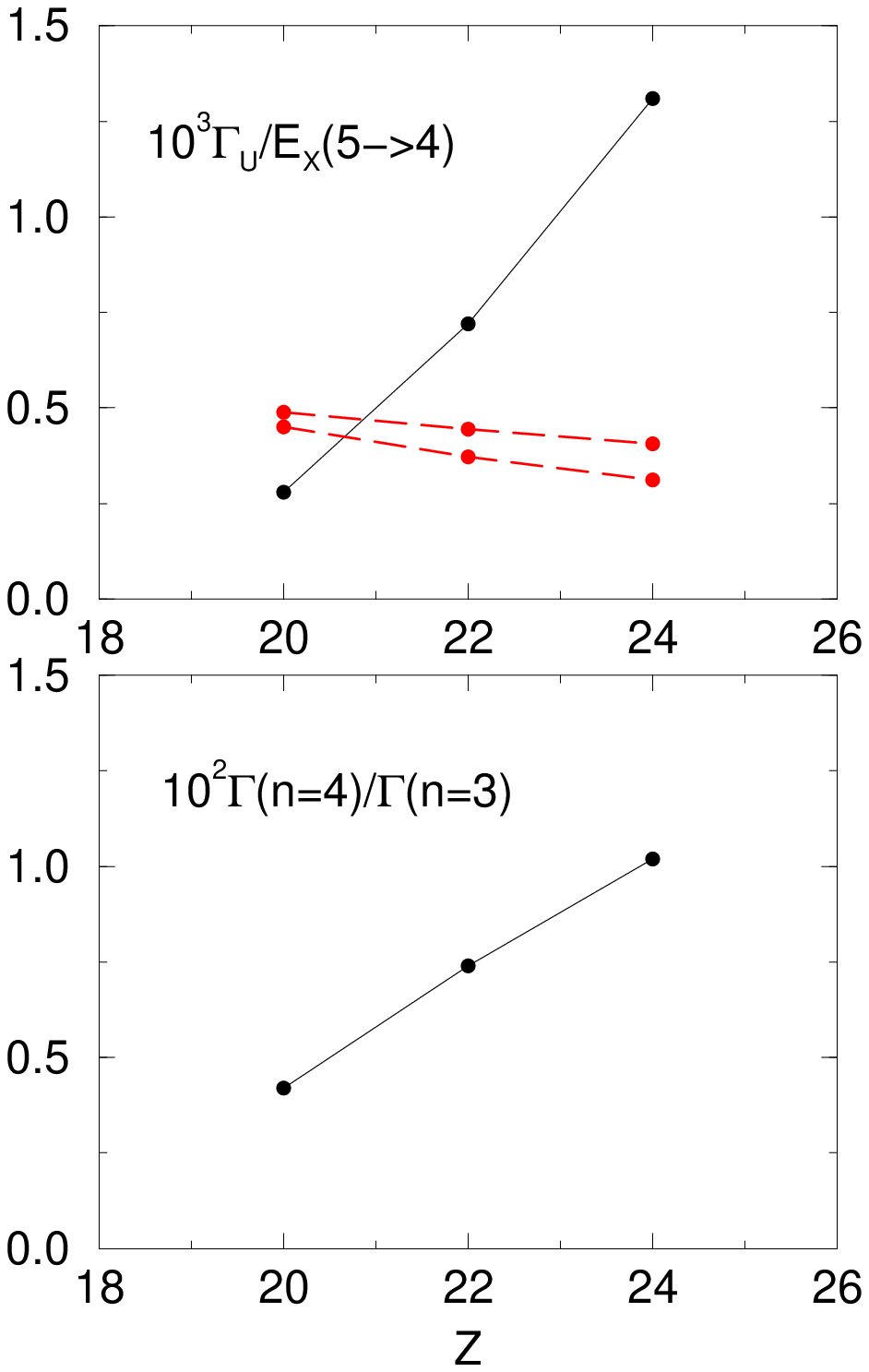}
\caption{Left: calculated strong-interaction widths of lower ($3d$) and upper
($4f$) levels and the X-ray
energy of the transition feeding into the upper level. Right: ratios of
upper to lower level widths (x10$^2$)
and ratios of the upper level width to the
energy of the transition into that level (x10$^3$).
The dashed lines represents estimated lower
limits for measuring $\Gamma _{\rm U}$, see text.}
\label{fig:43}
\end{center}
\end{figure}

We have calculated strong interaction width for the lower level 
($\Gamma _{\rm L}$) and for the upper level ($\Gamma _{\rm U}$) as well
as the energy E$_{\rm X}$ of the transition into the upper level 
in many kaonic atoms along the periodic table. Relative and absolute
yields of the relevant transition were also calculated.
In order to assess the feasibility of direct measurements 
of $\Gamma _{\rm U}$, 
this width should be compared with the expected 
resolution of the microcalorimeter
detector. Because of possible dependence on the measured energy itself, we
show in the following the relative resolution, i.e. 
$\Gamma _{\rm U}$/E$_{\rm X}$, the ratio of the width to the transition energy
to be measured. When this quantity is of the order of or larger than
the corresponding
ratio for the detector resolution then the experiment is deemed feasible. The 
various efficiencies are discussed later.

Starting with light elements, it became clear that the necessary conditions
for direct measurement of width of the upper levels in kaonic atoms
are hard to meet. Therefore we do not present any results for kaonic
atoms with 2$p$ as the lower level. This is not a severe restriction
considering that our interest in multinucleon absorption processes is
aimed at interactions in nuclear matter. 
 
\begin{figure}[htb]
\begin{center}
\includegraphics[height=95mm,width=0.45\textwidth]{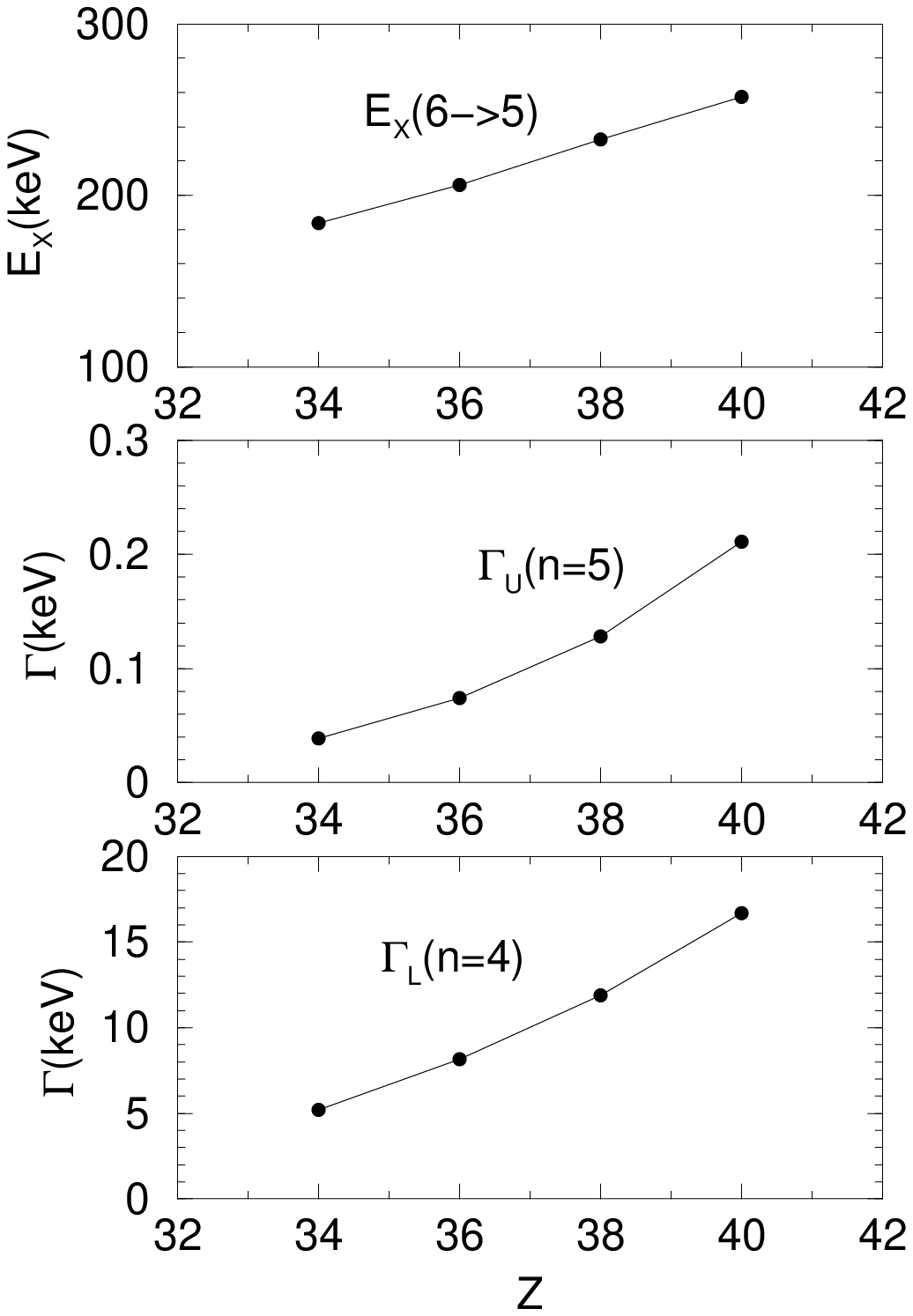}
\includegraphics[height=95mm,width=0.45\textwidth]{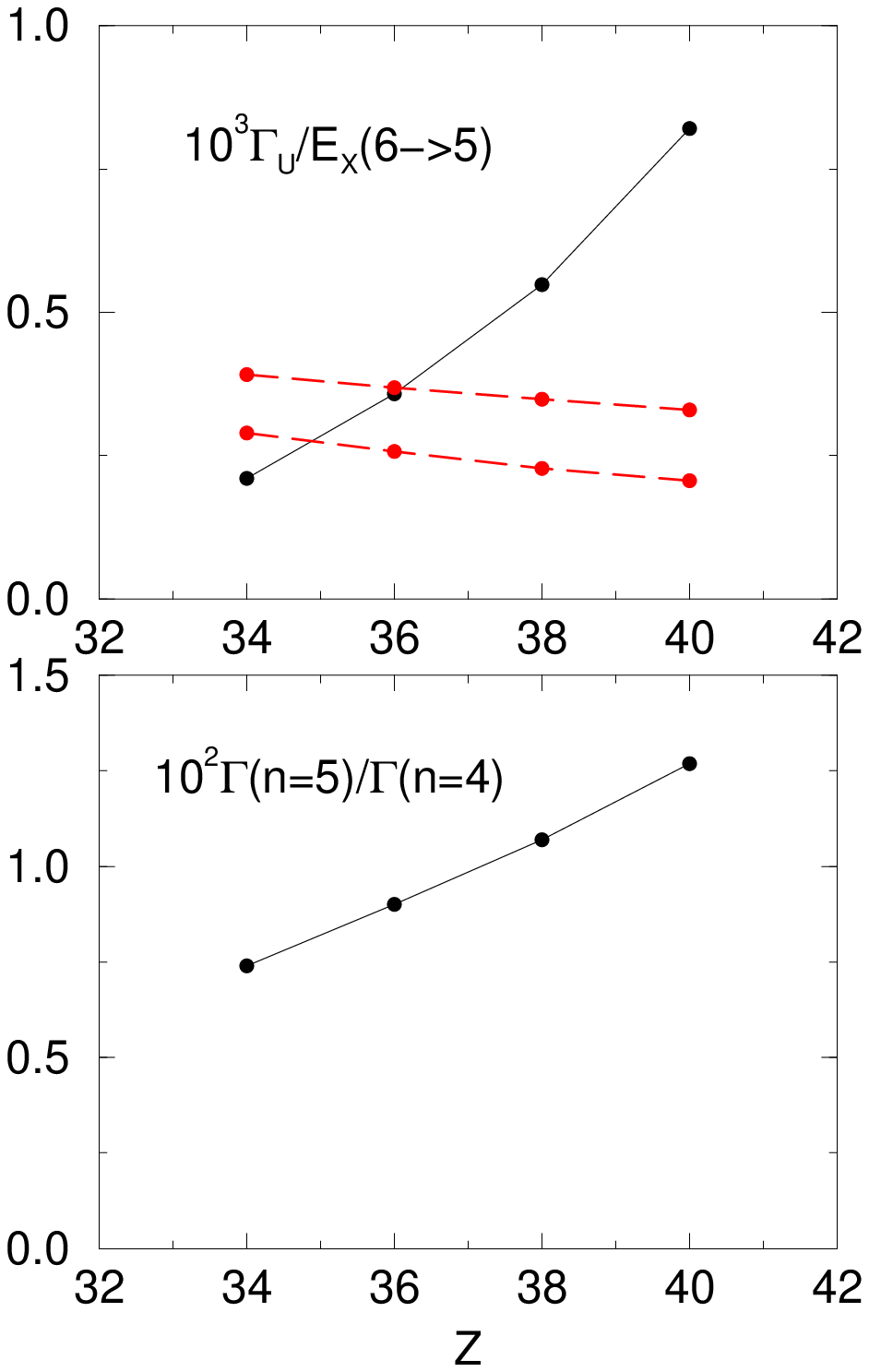}
\caption{Same as fig.\ref{fig:43} but for the $4f$ and $5g$ levels.}
\label{fig:54}
\end{center}
\end{figure}

\begin{figure}[htb]
\begin{center}
\includegraphics[height=95mm,width=0.45\textwidth]{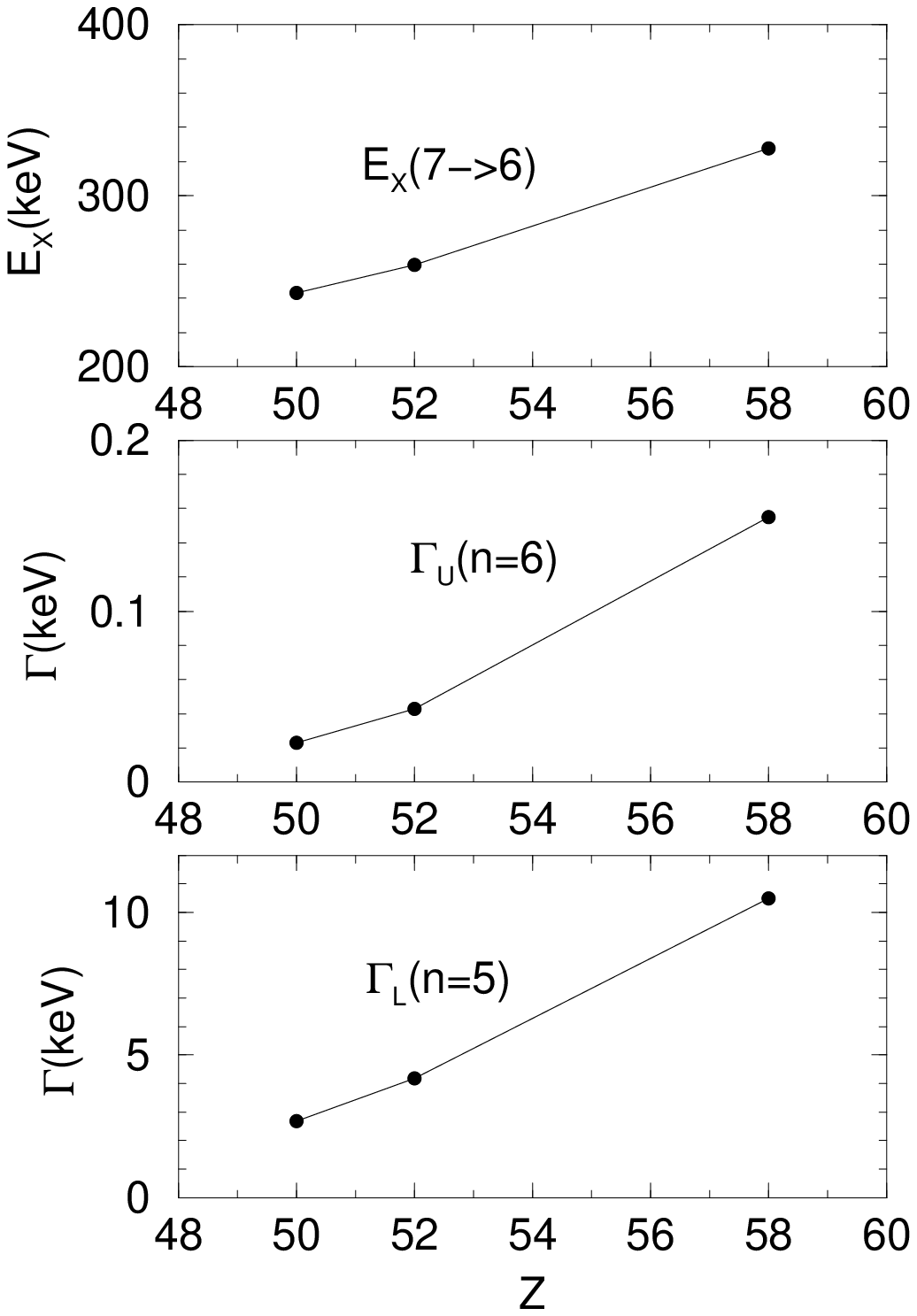}
\includegraphics[height=95mm,width=0.45\textwidth]{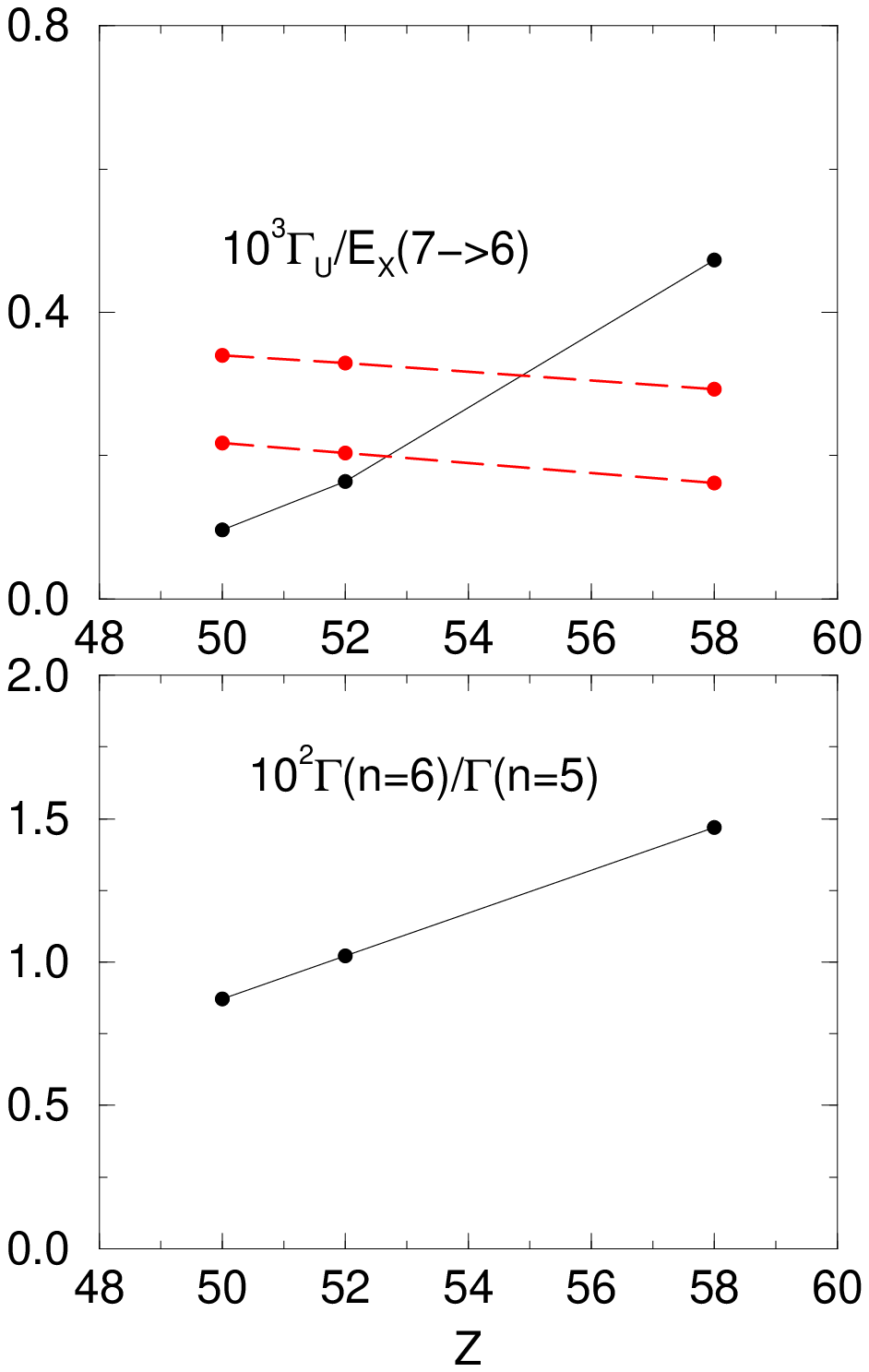}
\caption{Same as fig.\ref{fig:43} but for the $5g$ and $6h$ levels.}
\label{fig:65}
\end{center}
\end{figure}

Figure \ref{fig:43} shows results for a $3d$ lower level in 
kaonic atoms of Ca, Ti and Cr. The left-hand side shows widths of
the lower and of the upper level and also the energy of the X-ray
transition feeding into the upper level.
The right-hand side of the figure refers to the feasibility of 
a direct measurement of $\Gamma _{\rm U}$. The lower panel shows the
ratios $\Gamma _{\rm U}$/$\Gamma _{\rm L}$ (in units of 10$^{-2}$), 
to demonstrate that with
a conventional Ge detector which measures $\Gamma _{\rm L}$
it is usually unfeasible to measure also $\Gamma _{\rm U}$. The upper
panel refers to the direct measurement criterion where 
$\Gamma _{\rm U}$/E$_{\rm X}$ (continuous curve) should be 
comparable or larger
than the relative resolution of the detector (dashed curves) in order 
that the measurement be considered feasible.
The lower dashed curve represents a fixed value of $\Delta E$=53 eV
between 100 and 400 keV and the upper dashed curve is for 
$\Delta E = 53 \times \sqrt{E_X/100}$ eV where $E_X$ is in keV.
It is
seen that whereas Cr and possibly Ti could meet the feasibility 
criterion, the widths of the the lower levels are likely to be 
prohibitively large. Note  that all of the measured
lower level widths so far \cite{FGB94}
are well below 10~keV. Measuring larger widths 
could be adversely affected by background. It is evident that
there is some conflict between the requirements considering the two
widths.

Figure \ref{fig:54} shows similar results for the $4f$ lower and $5g$
upper levels. The conflict is here less severe, with Kr, Sr and Zr
exceeding the minimum criteria and with the width of the lower level
for Sr only slightly exceeding 10~keV. Further up the periodic table 
Fig. \ref{fig:65} shows results for the $5g$ lower and $6h$
upper levels and it is seen that if the $\Gamma_{\rm U}$/E$_{\rm X}$
criterion is relaxed somewhat, then for Z=54-56 the lower level
width is in the range of previous successful measurements.

\begin{figure}[htb]
\begin{center}
\includegraphics[height=95mm,width=0.45\textwidth]{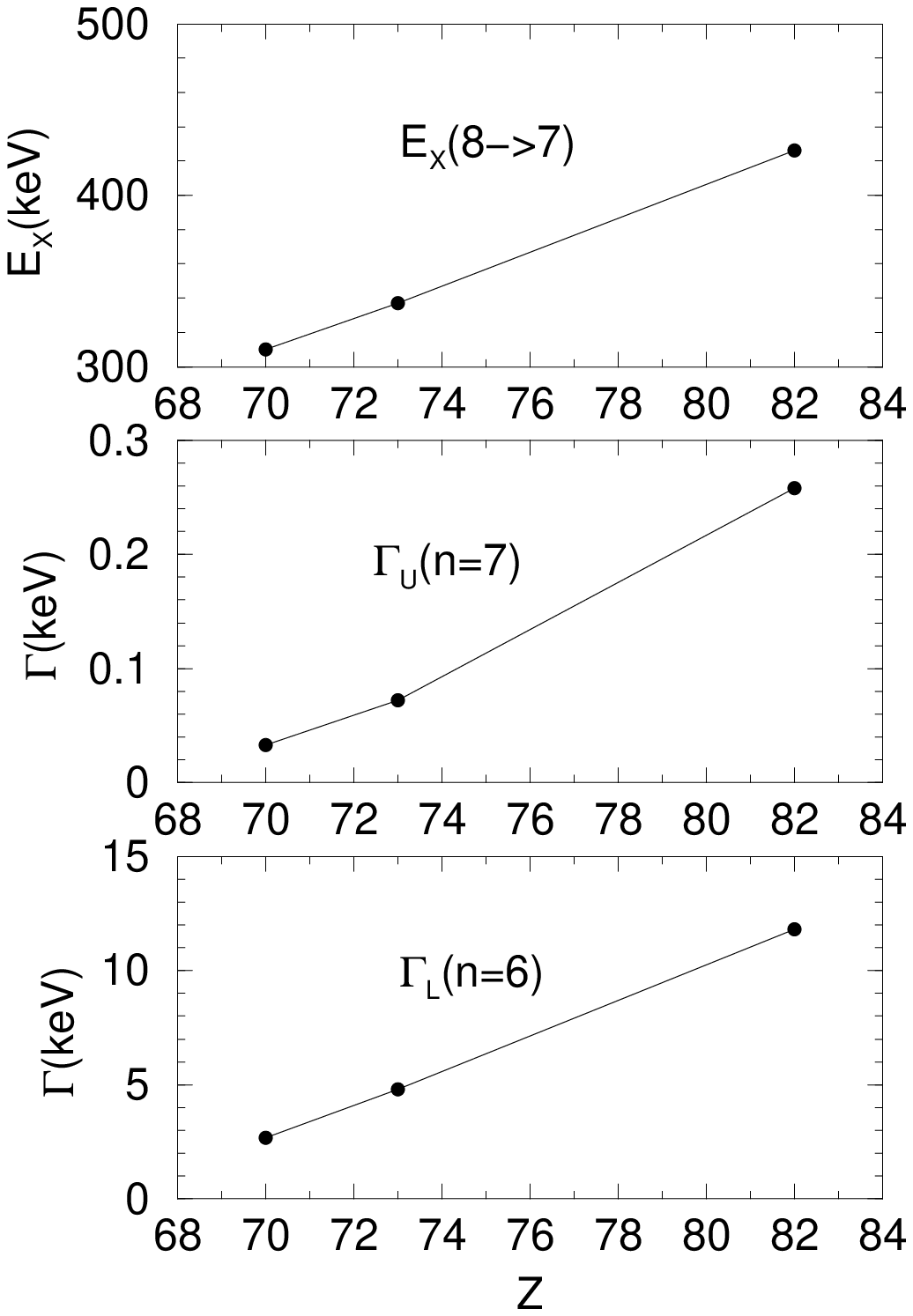}
\includegraphics[height=95mm,width=0.45\textwidth]{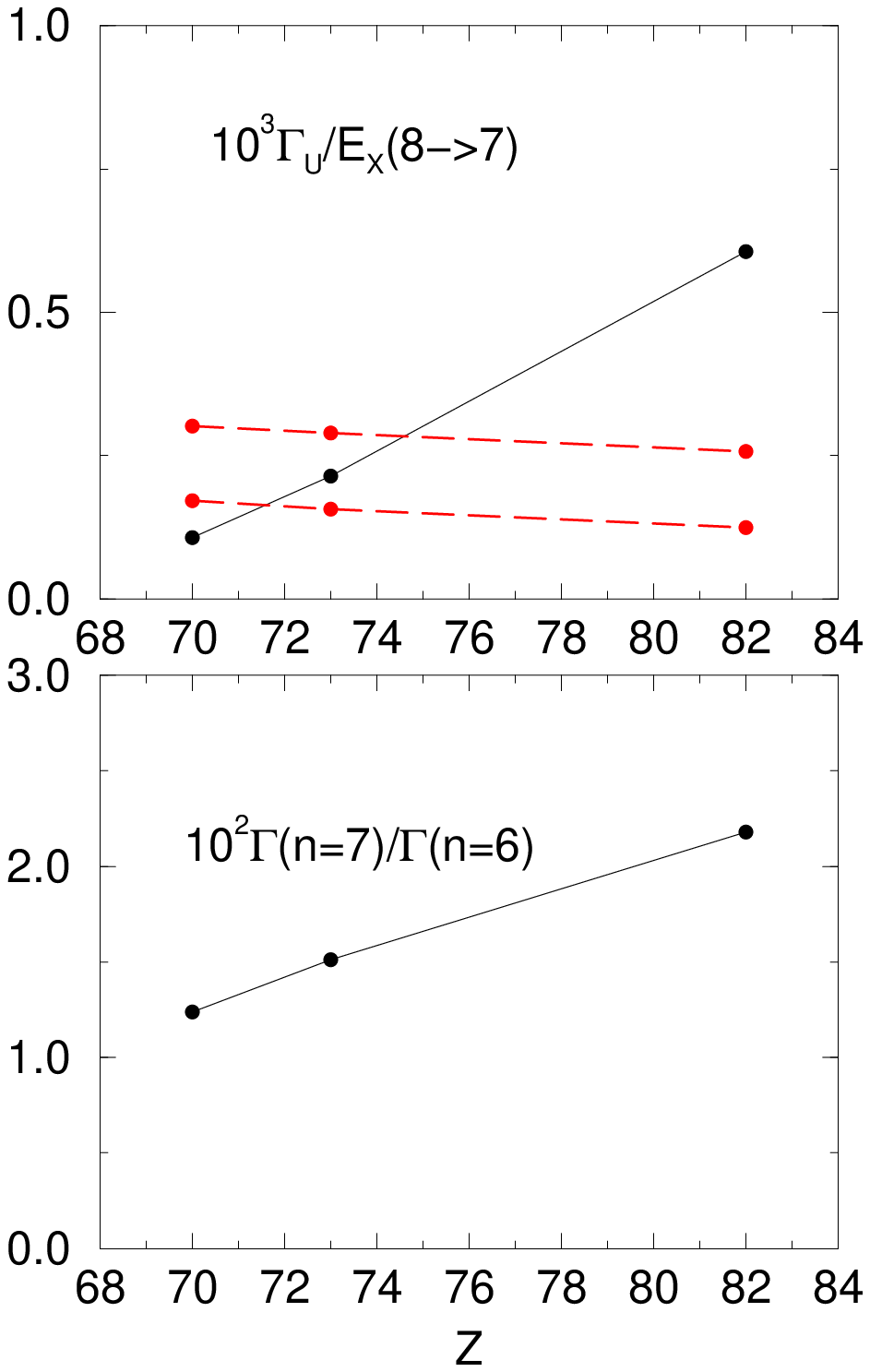}
\caption{Same as fig.\ref{fig:43} but for the $6h$ and $7i$ levels.}
\label{fig:76}
\end{center}
\end{figure}

Finally, Fig. \ref{fig:76} shows that heavy elements are the most suitable
targets for the required new kaonic atoms experiments. Note that strong 
interaction measurements in kaonic atoms of Yb, Ta and Pb have been
carried out before, albeit with rather modest accuracies.

Table \ref{tab:yields} summarizes values of relative and absolute
yields, calculated from the same potential as all other quantities
presented above. (The absolute yield is the X-ray yield per stopped
kaon).
 Relative yields are given for the upper to lower level
transitions. If upper level widths cannot be determined directly,
they can be determined from the measured relative yields,
 Eq. (\ref{eq:yield}). It is gratifying to note that all
the relative yields are smaller
than 0.25-0.33 so that the increase of the relative errors of the
corresponding width due to the 1/(1-Y$^{\rm rel}$) factor is not
too large.

Strong interaction widths of many other levels
are part of the input into the atomic cascade calculations 
and these were calculated from the same potential Eq. (\ref{eq:phen}).
The results in the table were calculated assuming the standard statistical
distribution of the population of $l$ values for large $n$. Varying
the parameter $\alpha$ in the range of $\pm$0.1 can change absolute
yields by up to $\pm$20\%. However, absolute yields do not serve
any purpose in the analysis; they should be regarded only
 as estimates that are useful in planning experiments.

\begin{table}
\caption{Absolute and relative yields of the relevant transitions
for kaonic atoms indicated in the figures. Also listed are
the ($n,l$) values of the various levels, see text.}
\label{tab:yields}
\begin{center}
\begin{tabular}{ccccccc}
\hline
target&U+1&U&L&Y$^{\rm abs}_{{\rm U}+1\rightarrow \rm U}$&
Y$^{\rm abs}_{{\rm U}\rightarrow \rm L}$
&Y$^{\rm rel}_{{\rm U}\rightarrow \rm L}$\\ \hline
Ca&(5,4)&(4,3)&(3,2)&0.665&0.044&0.061 \\
Ti&(5,4)&(4,3)&(3,2)&0.627&0.019&0.029 \\
Cr&(5,4)&(4,3)&(3,2)&0.570&0.012&0.019 \\
 & & & & & & \\
Se&(6,5)&(5,4)&(4,3)&0.705&0.091&0.121 \\
Kr&(6,5)&(5,4)&(4,3)&0.689&0.061&0.083 \\
Sr&(6,5)&(5,4)&(4,3)&0.663&0.043&0.062 \\
Zr&(6,5)&(5,4)&(4,3)&0.628&0.031&0.047 \\
 & & & & & & \\
Sn&(7,6)&(6,5)&(5,4)&0.706&0.218&0.292 \\
Te&(7,6)&(6,5)&(5,4)&0.696&0.152&0.207 \\
Ba&(7,6)&(6,5)&(5,4)&0.651&0.069&0.101 \\
 & & & & & & \\
Yb&(8,7)&(7,6)&(6,5)&0.674&0.236&0.331 \\
Ta&(8,7)&(7,6)&(6,5)&0.655&0.147&0.212 \\
Pb&(8,7)&(7,6)&(6,5)&0.601&0.067&0.107 \\ \hline
\end{tabular}
\end{center}
\end{table}

\begin{figure}[t]
 \begin{center}
  \includegraphics*[width=1.0\linewidth]{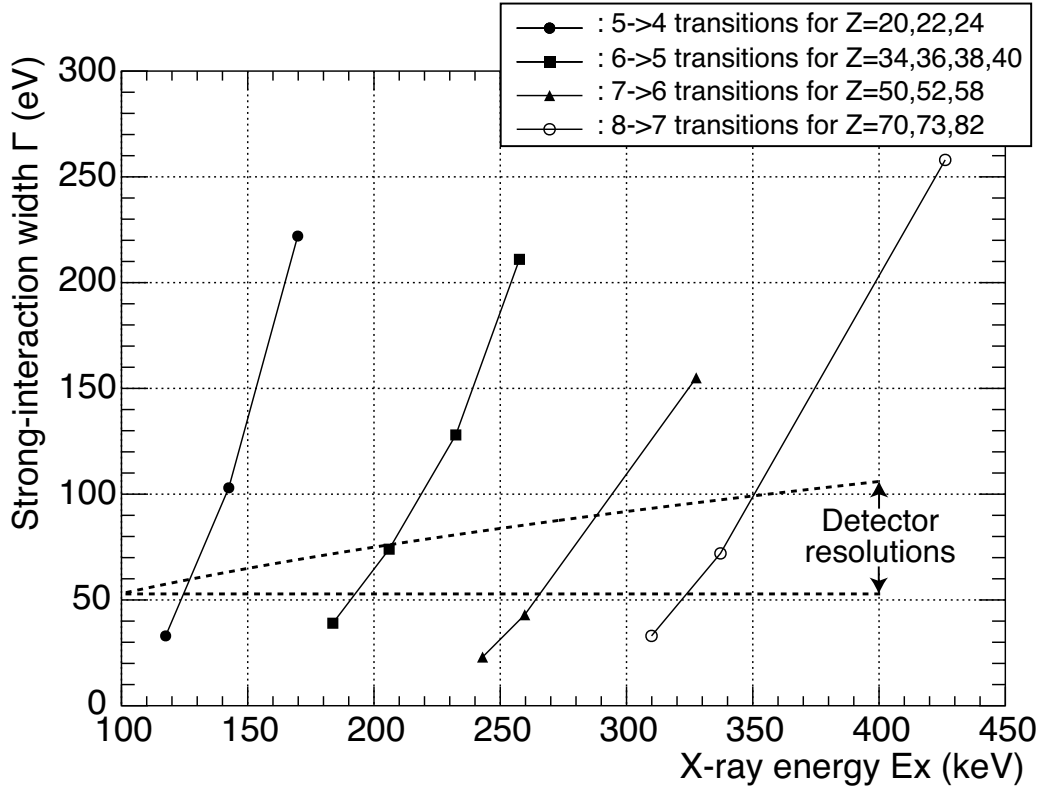} 
\caption{Summary of upper-level results and the feasibility guideline
due to the detector resolution.}
  \label{fig:WidthPlot}
 \end{center}
\end{figure}

Figure \ref{fig:WidthPlot} shows a summary of the present results
for upper levels and the guideline based on the detector resolution.
The lower dashed curve represents a fixed value of $\Delta E$=53 eV
between 100 and 400 keV and the upper dashed curve is for
$\Delta E = 53 \times \sqrt{E_X/100}$ eV where $E_X$ is in keV.
However, the widths of lower levels must also be taken into
consideration in planning experiments, as discussed above.
Almost all the
targets listed here have been used in previous exotic-atom
(pionic, kaonic and antiprotonic) experiments \cite{FGa07},
many as separated isotopes.

\section{Discussion and summary}
\label{sec:disc}

Along the atomic cascade in exotic atoms the radiation width
for the different levels varies rather slowly while the width for
absorption by the nucleus varies exponentially, as it depends on
the overlap of the atomic wavefunction with the nucleus, 
Eq. (\ref{eq:width}). That causes the cascade process to terminate
within one or two levels when the nuclear absorption dominates.
Direct measurements of strong interaction widths of two levels
require the measurement of two successive X-ray transitions.
The present results suggest that there could only be a handful
of elements where the required conditions are met and it raises 
the question of whether experiments on a few elements could match
the present data base which includes 24 different species of kaonic atoms.

The question of whether a partial set of data for kaonic atoms
 could provide similar
information to what is obtained from the full data was discussed
already in \cite{FGM99}, showing that indeed it was possible. 
Here we have compared the results of global fits to the 
present-day full data set
with results of similar fits to part of the data.
Selecting a partial set of the full data, namely 
Si, S, Ni, Cd, Sn, Yb and Ta that represent the four groups in
table \ref{tab:yields}, 
we obtain
parameter values and uncertainties very close to what is
found in fits to the full data. This suggests that a set of about six
targets, carefully selected, will be equivalent to the full
data base, particularly if the quality of the data is at
least as good as that of the old data.

Finally, test calculations show that uncertainties of
parameters of the best-fit  potentials indeed go down by close to
50\% if upper level widths replace relative yields in the data.
We conclude that new kaonic atom experiments on several carefully
selected targets using state-of-the-art microcalorimetric spectroscopy 
could provide widths of lower and upper levels
in the same atom which, in turn, could enhance our understanding
of multinucleon absorption of antikaons in the nuclear medium.

\hspace{25mm}

\section*{Acknowledgements}

Discussions with A.~Gal and correspondence with W.B. Doriese and J.N. Ullom 
are gratefully acknowledged.
This work was supported by the EU initiative FP7, 
HadronPhysics3, under the SPHERE and LEANNIS cooperation programs,
 by JSPS KAKENHI Grant Number 25105514 and by Incentive Research Grant
from RIKEN.

\hspace{30mm}

\end{document}